\documentclass[twocolumn,showpacs,preprintnumbers,amsmath,amssymb,floatfix]{revtex4}

\usepackage{graphicx}
\usepackage{dcolumn}
\usepackage{bm}

\usepackage[normalem]{ulem}  
\usepackage[dvips]{color}

\begin{document}


\title{The Statistical Multifragmentation Model with Skyrme Effective Interactions}

\author{S.R.\ Souza$^{1,2}$}
\author{B.V.\ Carlson$^{3}$}
\author{R.\ Donangelo$^{1,4}$}
\author{W.G.\ Lynch$^5$}
\author{A.W.\ Steiner$^5$}
\author{M.B.\ Tsang$^5$}
\affiliation{$^1$Instituto de F\'\i sica, Universidade Federal do Rio de Janeiro
Cidade Universit\'aria, \\CP 68528, 21941-972, Rio de Janeiro, Brazil}
\affiliation{$^2$Instituto de F\'\i sica, Universidade Federal do Rio Grande do Sul\\
Av. Bento Gon\c calves 9500, CP 15051, 91501-970, Porto Alegre, Brazil}
\affiliation{$^3$Departamento de F\'\i sica, 
Instituto Tecnol\'ogico de Aeron\'autica - CTA, 12228-900\\
S\~ao Jos\'e dos Campos, Brazil}
\affiliation{$^4$Instituto de F\'\i sica, 
Facultad de Ingenier\'\i a, Universidad de la Rep\'ublica,\\
Julio Herrera y Reissig 565, 11.300 Montevideo, Uruguay}
\affiliation{$^5$ Joint Institute for Nuclear Astrophysics, National
Superconducting Cyclotron Laboratory, and the Department of Physics
and Astronomy, Michigan State University,
East Lansing, MI 48824, USA}

\date{\today}

\begin{abstract}
The Statistical Multifragmentation Model is modified to incorporate
the Helmholtz free energies calculated in the finite temperature Thomas-Fermi
approximation using Skyrme effective interactions.
In this formulation, the density of the fragments at the freeze-out 
configuration corresponds to the equilibrium value obtained in the 
Thomas-Fermi approximation at the given temperature.
The behavior of the nuclear caloric curve at constant volume is investigated
in the micro-canonical ensemble and a plateau is observed for excitation
energies between 8 and 10 MeV per nucleon.
A kink in the caloric curve is found at the onset of this gas transition,
indicating the existence of a small excitation energy region with
negative heat capacity.
In contrast to previous statistical calculations, this situation takes 
place even in this case in which the system is constrained to fixed volume.
The observed phase transition takes place at approximately constant entropy.
The charge distribution and other observables also turn out to be sensitive
to the treatment employed in the calculation of the free energies and the 
fragments' volumes at finite temperature, specially at high excitation energies.
The isotopic distribution is also affected by this treatment, which suggests
that this prescription may help to obtain information on the
nuclear equation of state.
\end{abstract}

\maketitle

\section{\label{sec:introuduction}Introduction\protect}
Understanding the behavior of nuclear matter far from equilibrium,
besides its intrinsic relevance to theoretical nuclear physics, is a subject
of great interest to nuclear astrophysics, where the fate of supernovae or the 
properties of neutron stars are appreciably influenced by the nuclear equation 
of state (EOS) \cite{Bethe1990,EOSNeutronStar,EOS2}.
Thus, this area has been intensively investigated in different contexts during 
the last decades \cite{EOSNeutronStar,EOS2,EOS1,EOS3,EOS4,EOS5,EOS6,BillScience,
EOSLi,PawelFlow}.
Nuclear collisions, at energies starting at a few tens of MeV per nucleon, 
provide a suitable means to study hot and compressed nuclear matter
\cite{BillScience,EOSLi,PawelFlow,exoticDens,AichelinPhysRep,bkDensViola,
XLargeSystems,FlowSchussler,ccNatowitz}.
The determination of the nuclear caloric curve is of particular interest as it
allows one to infer on the existence of a liquid-gas phase transition in nuclear
matter.
Nevertheless, owing to experimental difficulties, conflicting observations have 
been made in different experimental analyses
\cite{ccNatowitz,ccMa2005,ccDagostino,ccSharenberg,ccCampi,ccRuangma,
reviewSubal2001,ccKwiatkowski,ccSerfling,ccXi,ccAuger,ccMa1997,ccMoretto1996,ccGSI}.
Although there have been attempts to reconcile these results 
\cite{ccNatowitzHarm}, this issue has not been settled.

The properties of the disassembling system in central collisions, as well as the
outcome of the reactions, have been found to be fairly sensitive to the EOS 
employed in the many theoretical studies using dynamical models 
that have been performed
\cite{BillScience,AichelinPhysRep,EOSLi,PawelFlow,exoticDens}.
However, in spite of their success in describing many features of the nuclear 
Multifragmentation process 
\cite{Bondorf1995,BettyPhysRep2005,GrossMMM1990}, 
there has not been much effort to incorporate information based on the EOS 
in the main ingredients of statistical multifragmentation models.
Yet, these models have recently been applied to investigate, for instance, the
Isospin dependence of the nuclear energy at densities below the saturation value 
\cite{isoscalingIndraGSI2005,isoSymmetryBotvina2006,symEnergyShatty2007}.
These calculations have suggested an appreciable reduction of the symmetry 
energy coefficient at low densities but other statistical calculations
\cite{RadutaIsoSym1,RadutaIsoSym2,isoMassFormula2008} indicate that surface 
corrections to the symmetry energy may also explain the behavior observed in
those studies.
Therefore, statistical treatments, which consistently include density effects, 
are most advisable for these studies.

In this work, we modify the Statistical Multifragmentation Model (SMM)
\cite{smm1,smm2,smm4} and calculate some of its key ingredients from the finite 
temperature Thomas-Fermi approximation
\cite{TFBrack1,TFBrack2,SuraudVautherin,Suraud1987} using Skyrme effective interactions.
This version of the model is henceforth labeled SMM-TF.
The internal Helmholtz free energies of the fragments are
calculated in a mean field approximation, which is fairly sensitive 
to the Skyrme force used \cite{BLV2}. 
This makes possible to investigate whether such statistical treatments may 
provide information on the EOS.
Furthermore, this approach allows to consistently take into account 
contributions to the free energy due to excitations in the continuum, in 
contrast to the traditional SMM \cite{ISMMlong}.
For consistency with the mean field treatment, the equilibrium density of the 
fragments at the freeze-out stage is also provided by the Thomas-Fermi 
calculations.
Thus, in contrast with former SMM calculations, fragments are allowed to be 
formed at densities below their saturation value.
For a fixed freeze-out volume, this leads to a systematic reduction of the free 
volume, which directly affects the entropy of the fragmenting system, 
the fragment's kinetic energies, and, also, the system's pressure.
As a consequence, other properties, such as the caloric curve and the 
multiplicities of the different fragment species produced, are also 
affected.

We have organized the remainder of this work as follows.
In Sect.\ \ref{sec:model} we discuss the modifications to the SMM and
present the results obtained with this modified treatment in 
Sect.\ \ref{sec:results}.
Concluding remarks are drawn in Sect.\ \ref{sec:conclusions}.
In Appendix \ref{sec:ThomasFermi} we provide a brief description of the 
Thomas-Fermi calculations employed in this work.

\section{\label{sec:model}Theoretical framework\protect}

In the SMM \cite{smm1,smm2,smm4}, it is assumed that a source, made up
of $Z_0$ protons and $A_0-Z_0$ neutrons, is formed at the late stages of the
reaction, with total excitation energy $E^*$.
This excited source then undergoes a simultaneous statistical breakup.
As the system expands, there is a fast exchange of particles among the 
different fragments until a freeze-out configuration is reached, at which 
time particle exchange ceases and the composition of the primary fragments 
is well defined.
One then assumes that thermal equilibrium has been reached and calculates the 
properties of the possible fragmentation modes through the laws of equilibrium 
statistical mechanics.
A possible scenario consists in conjecturing that the freeze-out configuration 
is always attained when the system reaches a fixed pressure, {\it i.e.} the 
nuclear multifragmentation is an isobaric process.
In this case, different statistical calculations predict a plateau in the 
caloric curve 
\cite{smm2,negHC1,negHC2,negHC3,GrossPhysRep,negHC4,smmIsobaric}.
The situation is qualitatively different if one assumes that, for a given 
source, the freeze-out configuration is reached at a fixed breakup volume 
$V_\chi$.
As found in many different calculations, a monotonic increase of the 
temperature with excitation energy takes place in this case
\cite{isocc,ccBotvina,smmIsobaric}. 
In what follows we demonstrate that this is a consequence of the properties
assumed for the fragments formed, and not of the fixed volume assumption.

In this work we keep the breakup volume fixed for all fragmentation 
modes, and parametrize it through the expression:

\begin{equation}
V_\chi=(1+\chi)V_0\;,
\label{eq:vbk}
\end{equation}

\noindent
where $V_0$ denotes the volume of the system at normal density and $\chi\ge 0$
is an input parameter.

In the micro-canonical version of SMM, the sampled fragmentation modes 
\cite{smm4} are consistent with mass, charge, and energy conservation and thus 
the following constraints are imposed for each partition:

\begin{equation}
A_0=\sum_{A,Z}N_{A,Z} A\;,
\label{eq:a0}
\end{equation}

\begin{equation}
Z_0=\sum_{A,Z}N_{A,Z} Z\;,
\label{eq:z0}
\end{equation}

\noindent
and

\begin{eqnarray}
&&E_{\rm source}^{\rm g.s.}+E^* = E_{\rm trans}(T)
+\sum_{A,Z}N_{A,Z}\left[-B_{A,Z}+\epsilon^{*}_{A,Z}\right]\nonumber\\
&&+ \frac{C_{\rm Coul}}{(1+\chi)^{1/3}}\frac{Z_0^2}{A_0^{1/3}}
-\frac{C_{\rm Coul}}{(1+\chi)^{1/3}}\sum_{A,Z}N_{A,Z}\frac{Z^2}{A^{1/3}}\;.
\label{eq:econs}
\end{eqnarray}

\noindent
In the above equations, $E_{\rm source}^{g.s.}$ is the ground state energy of 
the source, $N_{A,Z}$ denotes the multiplicity of fragments with mass 
and atomic numbers $A$ and $Z$, respectively, 
$B_{A,Z}$ corresponds to the binding energy of the fragment, and
$\epsilon^*_{A,Z}(T)$ represents its excitation energy at temperature $T$.
The Coulomb repulsion among the fragments is taken into account by the last 
two terms of the above equation which, together with the self energy 
contribution included in $B_{A,Z}$, give the Wigner-Seitz \cite{WignerSeitz} 
approximation discussed in Ref.\ \cite{smm1}.
The coefficient $C_{\rm Coul}$ is given in Ref.\ \cite{ISMMmass}.
As discussed in Ref.\ \cite{ISMMlong}, the fragment's binding energy $B_{A,Z}$ 
is either taken from experimental values \cite{AudiWapstra} or is obtained from 
a careful extrapolation if empirical information is not available.
The spin degeneracy factors, which enter in the calculation of the translational 
energy $E_{\rm trans}$, are also taken from experimental data for $A\le 4$.
In the case of heavier fragments, this factor is neglected, {\it i.e.}, it is
set to unity for all nuclei.

One should notice that the freeze-out temperature varies from one fragmentation 
mode $f=\{N_{A,Z}\}$ to another, since it is determined from the 
energy conservation constraint of Eq.\ (\ref{eq:econs}).
Therefore, the average temperature is calculated, as any other observable $O$, 
through the usual statistical averages:

\begin{equation}
\langle O\rangle = \frac{\sum_f O_f\exp(S_f)}{\sum_f\exp(S_f)}\;,
\label{eq:observ}
\end{equation}

\noindent
where $S_f$ denotes the entropy associated with the mode $f$.
This entropy is calculated through the standard thermodynamical relation

\begin{equation}
S=-\frac{dF}{dT}\;,
\label{eq:sstd}
\end{equation}

\noindent
where

\begin{equation}
F=E-TS
\label{eq:fstd}
\end{equation}

\noindent
is the Helmholtz free energy.
In the following, we write this quantity as

\begin{equation}
F=\sum_{A,Z}N_{A,Z}\left[-B_{A,Z}+f^*_{A,Z}(T)+f^{\rm trans}_{A,Z}(T)\right]+F_{\rm Coul}
\label{eq:fesplit}
\end{equation}

\noindent
where the contributions from the fragment's internal excitation ($f^*_{A,Z}$) 
and translational motion ($f^{\rm trans}_{A,Z}$) are explicitly separated.
The latter reads:

\begin{equation}
f_{A,Z}^{\rm trans}=-T\left[\log\left(\frac{g_{A,Z}V_fA^{3/2}}{\lambda_T^3}
                    \right)-\frac{\log(N_{A,Z}!)}{N_{A,Z}}\right]\;.
\label{eq:faz}
\end{equation}

\noindent
In the above expression, $\lambda_T=\sqrt{\frac{2\pi\hbar^2}{m_nAT}}$ 
is the thermal wavelength, $m_n$ is the nucleon mass, $g_{A,Z}$ is 
the spin degeneracy factor, and $V_f$ denotes the free volume, {\it i.e.}, it is 
the difference between $V_\chi$ and the volume occupied by all the fragments 
at freeze-out.
The quantity $F_{\rm Coul}$ corresponds to the last two terms in 
Eq.\ (\ref{eq:econs}).

Before we present the changes in the model associated with the Thomas-Fermi 
calculations, we briefly recall below the calculation of Helmholtz free energy
$F$ in the SMM.

\subsection{\label{sec:ISMM}The standard SMM}
In its original formulation \cite{smm1}, the SMM assumes that the diluted 
nuclear system undergoes a prompt breakup and that the 
resulting pieces of matter collapse to normal nuclear density, although 
being at temperature $T$.
Therefore, the volume occupied by the fragments corresponds to $V_0$, so that

\begin{equation}
V_f=\chi V_0\;.
\label{eq:vfstd}
\end{equation}

\noindent
The energy and entropy associated with the translational motion of the fragment 
are respectively given by:

\begin{equation}
\epsilon^{\rm trans}_{A,Z}=f^{\rm trans}_{A,Z}+Ts^{\rm trans}_{A,Z}=\frac{3}{2}T
\label{eq:etransstd}
\end{equation}

\noindent
and

\begin{eqnarray}
s^{\rm trans}_{A,Z}&=&-\frac{d}{dT}f^{\rm trans}_{A,Z}\nonumber\\
                   &=&  \frac{3}{2}+\log\left[
                   \frac{g_{A,Z}V_fA^{3/2}}{\lambda_T^3}\right]
                    -\frac{\log(N_{A,Z}!)}{N_{A,Z}}\;.
\label{eq:stransstd}
\end{eqnarray}

The internal free energy $f^*_{A,Z}$ has contributions from bulk
and surface terms:

\begin{equation}
f^*_{A,Z}=-\frac{T^2}{\epsilon_0}A+\beta_0A^{2/3}
\left[\left(\frac{T_c^2-T^2}{T_c^2+T^2}\right)^{5/4} -1\right]\;.
\label{eq:feint0}
\end{equation}

\noindent
The values of the parameters in the above expression are 
$\epsilon_0=16.0$~MeV, $T_c=18.0$~MeV and $\beta_0=18.0$~MeV \cite{ISMMlong}.
This expression is used for all nuclei with $A\ge 5$.
Lighter fragments are assumed to behave as point particles, except for the alpha particle,
for which one retains the bulk contribution to the free energy in order to
take its excited states into account.

In Ref.\ \cite{ISMMlong}, the calculation of $f^*_{A,Z}$ has been modified to
include empirical information on the excited states of light nuclei.
We label this version of the model as ISMM and it is used throughout this work.

\subsection{\label{sec:SMMTF}The SMM-TF}
\begin{figure}[tb]
\includegraphics[height=6.0cm]{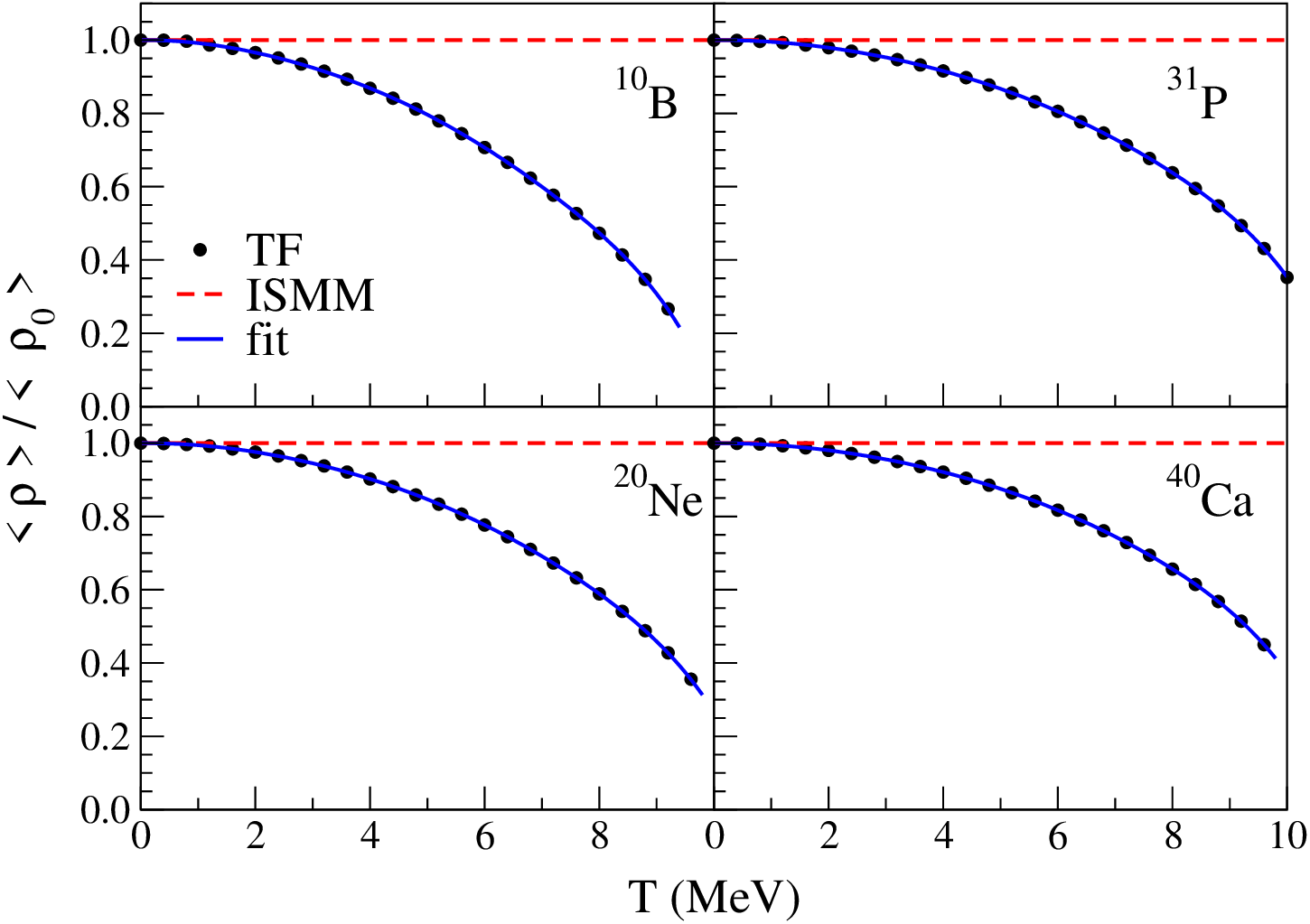}
\caption{\label{fig:aveRho} (Color online) Ratio between the average equilibrium 
density of the nucleus at temperature $T$ and the ground state value as a function 
of the temperature. For details, see the text.
}
\end{figure}

The Thomas-Fermi approximation, briefly outlined in Appendix \ref{sec:ThomasFermi}, 
allows one to calculate the internal free energy of the fragments $f^*_{A,Z}$ from 
Skyrme effective interactions.
Equations (\ref{eq:fengg},\ref{eq:feaztf}) clearly show that $f^*_{A,Z}$ 
contains, besides those from the nuclear interaction traditionally used in SMM, 
contributions associated with the Coulomb energy in addition to the ones appearing 
in [Eq.\ (\ref{eq:feint0})].
The additional Coulomb contribution arises, in the present case, because the 
equilibrium density of the nucleus at temperature $T$ does not correspond, in general, 
to its ground state value.
This is illustrated in Fig.\ \ref{fig:aveRho}, which shows the ratio between 
the average density $\langle\rho\rangle$ at a temperature $T$, and the corresponding 
ground state value $\langle\rho_0\rangle$, for several selected light 
and intermediate mass nuclei.
We define the sharp cutoff density $\langle\rho\rangle$ as that which gives the same 
root mean square radius as the actual nuclear density obtained in the Thomas-Fermi 
calculation.
One observes that $\langle\rho\rangle$ decreases as one increases the temperature of 
the nucleus and that it quickly goes to zero as $T$ approaches its limiting temperature,
since the nuclear matter tends to move to the external border of the box due to
Coulomb instabilities \cite{BLV1,BLV2,Suraud1987}.
In our SMM-TF calculations presented below, we only accept a fragmentation mode 
at temperature $T$ if it is smaller than the limiting temperature of all the fragments 
of the partition.
If this is not the case, the entire partition is discarded as not being physically
possible and we sample another one.

Thus, a fragment's volume at temperature $T$ is defined as:

\begin{equation}
\frac{V_{A,Z}}{V_{A,Z}^0}=\frac{\langle\rho_0^{A,Z}\rangle}{\langle\rho^{A,Z}\rangle}\;,
\end{equation}

\noindent
where $V_{A,Z}^0$ represents the ground state value.
Since it is useful to have analytical formulae to use in practical SMM calculations, 
we performed a fit of $\langle\rho^{A,Z}\rangle$ using the following expression

\begin{equation}
\frac{\langle\rho^{A,Z}\rangle}{\langle\rho_0^{A,Z}\rangle}=1+T^{a_n^{A,Z}}
\sum_{i=0}^{n-1}a_i^{A,Z}T^i\;,
\label{eq:fitRho}
\end{equation}

\noindent
where $\{a_i^{A,Z}\}$ are the fit parameters.
This expression has proven to be accurate enough for numerical applications, 
as is illustrated in Fig.\ \ref{fig:aveRho}, which shows a comparison between 
Eq.\ (\ref{eq:fitRho}) (full lines) and the results obtained with the Thomas-Fermi 
calculation (circles).
The fit was carried out using $n=6$.
The dashed lines emphasize the fact that 
$\langle\rho^{A,Z}\rangle=\langle\rho^{A,Z}_0\rangle$
in the standard SMM.

Instead of being given by Eq.\ (\ref{eq:vfstd}), the free volume of a fragmentation
mode now reads

\begin{equation}
V_f=(1+\chi)V_0-\sum_{A,Z}N_{A,Z}V^0_{A,Z}
\left[1+T^{a_n^{A,Z}}\sum_{i=0}^{n-1}a_i^{A,Z}T^i\right]^{-1}\;.
\label{eq:vftf}
\end{equation}

\noindent
For the values of $\chi$ usually adopted in statistical calculations
($0\le\chi\le 5$), this expression shows that, for some partitions,
there may be a temperature $T_V$ for which $V_f \le 0$.
Therefore, if Eq.\ (\ref{eq:econs}) leads to $T\ge T_V$,
the partition is discarded as it is not a physically acceptable solution. 

From Eq.\ (\ref{eq:faz}), the entropy associated with the kinetic motion of the fragment
$(A,Z)$ becomes

\begin{equation}
s^{\rm trans}_{A,Z}=  \frac{3}{2}+\log\left[\frac{g_{A,Z}V_f A^{3/2}}
                      {\lambda_T^3}\right]
                   -\frac{\log(N_{A,Z}!)}{N_{A,Z}}+\frac{T}{V_f}\frac{dV_f}{dT}\;.
\label{eq:stransfe}
\end{equation}

\begin{figure}[tb]
\includegraphics[height=6.0cm]{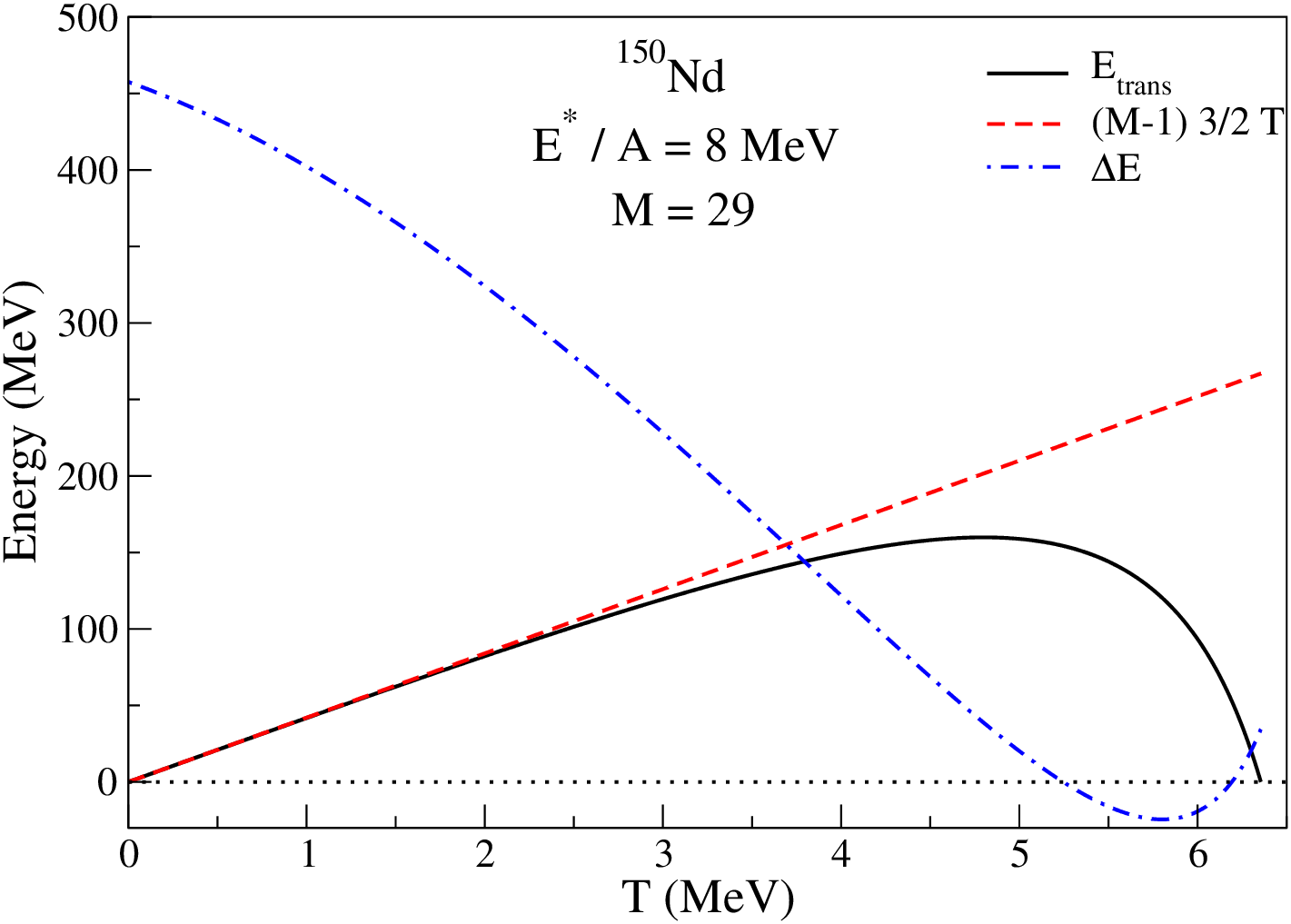}
\caption{\label{fig:eZero} (Color online) Kinetic energy (full line) of a 
particular partition of the $^{150}$Nd nucleus into $M=29$ fragments, 
for $E^*/A = 8$~MeV and $V_\chi/V_0=3$, as a function of the temperature.
For comparison, the standard average translational energy, $(M-1)\frac{3}{2}T$, is also
displayed (dashed line).
The difference between the left and the right hand sides of Eq.\ (\ref{eq:econs}),
$\Delta E$, is also shown (dashed-dotted line).
}
\end{figure}

\noindent
One should notice that, besides the smaller free volume, the last term in the above 
expression does not appear in the earlier version of the SMM.
Since $dV_f/dT \le 0$, the expression above gives a smaller contribution 
to the total entropy than Eq.\ (\ref{eq:stransstd}).
Owing to this change in the entropy, the average kinetic energy of the fragment
becomes:

\begin{equation}
\epsilon_{A,Z}^{\rm trans}=\frac{3}{2}T\left(1+\frac{2}{3}\frac{T}{V_f}
                           \frac{\partial V_f}{\partial T}\right)\;,
\label{eq:etranstf}
\end{equation}

\noindent
which, for a given temperature, is also lower than the corresponding SMM value.
As a matter of fact,  if the second factor dominates the first one, for
$T>T_K$, where $\epsilon^{\rm trans}_{A,Z}(T_K)=0$, it can even become negative.
We also discard all partitions for which there is no solution of Eq.\ (\ref{eq:econs}) 
satisfying $T<T_K$.
This aspect is illustrated in Fig.\ \ref{fig:eZero}, which shows the total kinetic 
energy $E_{\rm trans}$ as a function of $T$ for the $^{150}$Nd nucleus, with
$E^*/A = 8$~MeV, for a partition containing $M=29$ fragments.
The full line represents $E_{\rm trans}$, whereas the dashed line corresponds to 
the standard SMM formula.
The factor, $M-1$ is due to the fact that the center of mass motion is consistently 
removed in all the kinetic formulae, although it is not explicitly stated above.

The observed drop of the kinetic energy may lead to nontrivial consequences.
In the case of the $^{150}$Nd nucleus and for $E^*/A\lesssim 7.0$~MeV, the 
fragment multiplicity is relatively low.
Therefore, in this lower excitation energy range the behavior 
of the kinetic energy does not lead to any qualitative changes arising from
the energy conservation constraint.
However, for higher excitation energies, and consequently larger fragment multiplicities, 
the kinetic energy is comparable to the total energy of the system $E_{\rm total}$.
In this case, for a given value of $E_{\rm total}$, there may be two values of $T$ which
are acceptable solutions to Eq.\ (\ref{eq:econs}).
This is also illustrated in Fig.\ \ref{fig:eZero}, which shows the difference $\Delta E$ 
between the left and the right hand sides of this expression.
Since all the micro-states corresponding to the same total energy $E_{\rm total}$ should 
be included, both solutions, in this case associated with temperatures
$T\approx 5.3$~MeV and $T\approx 6.2$~MeV must be considered.
They contribute, however, with different statistical weights, due to the different number 
of states associated with each of these two solutions.

\begin{figure}[tb]
\includegraphics[height=6.0cm]{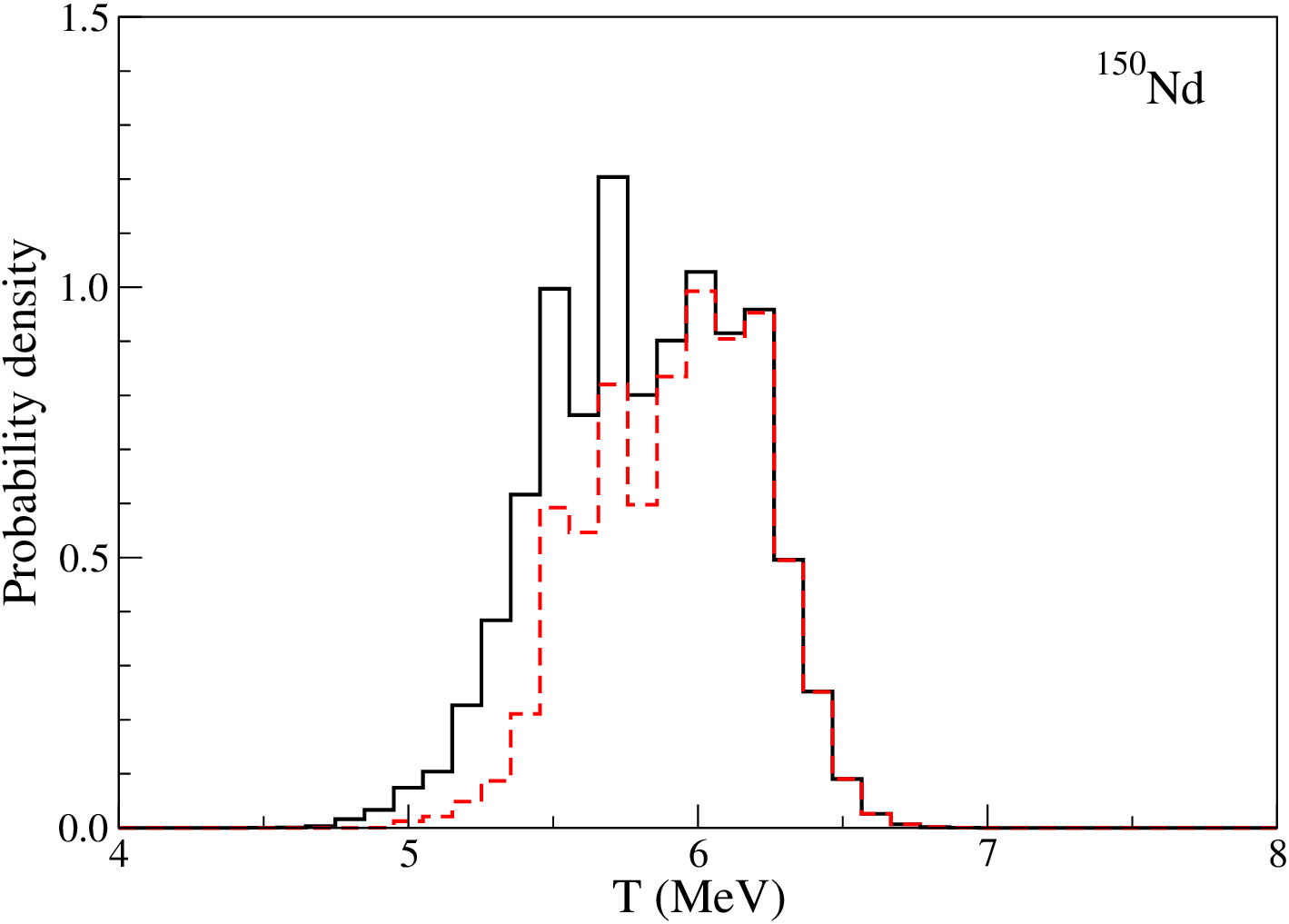}
\caption{\label{fig:histT} (Color online) Temperature distribution for the breakup of the
the $^{150}$Nd nucleus at $E^*/A = 8$~MeV and for $V_\chi/V_0=3$.
The full line corresponds to all events while those in which there are two temperatures
associated with $E_{\rm total}$ are depicted by the dashed line.
}
\end{figure}

Based on this scenario, the determination of the freeze-out temperature from isotopic 
ratios \cite{Albergo}, where one tacitly assumes that $T$ is univocally determined from $E^*$, 
should be carefully reexamined.
To give a quantitative estimate of these effects, we show,
in Fig.\ \ref{fig:histT}, the temperature distribution for the fragmentation of 
the $^{150}$Nd nucleus, at $E^*/A=8$~MeV and $V_\chi=3V_0$.
The full line in this picture shows the results when all the partitions are considered
whereas the dashed line represents only those which lead to two different temperatures.
The cases where there are two temperature solutions correspond to 43\% of the events 
and account for 76\% of the total statistical weight.
These numbers are drastically changed at lower excitation energies where, for instance, 
one finds, at $E^*/A=6$~MeV, 0.03\% and 0.09\%, respectively.
In spite of the great importance of these solutions at high excitation energy, the temperature
distribution does not exhibit two clear dominant peaks, separated by a gap, as it could be 
expected from Fig.\ \ref{fig:eZero}.
This is because the numerical value of the two solutions vary from one
partition to the other and the expected signature is thus blurred.

\begin{figure}[tb]
\includegraphics[height=5.5cm]{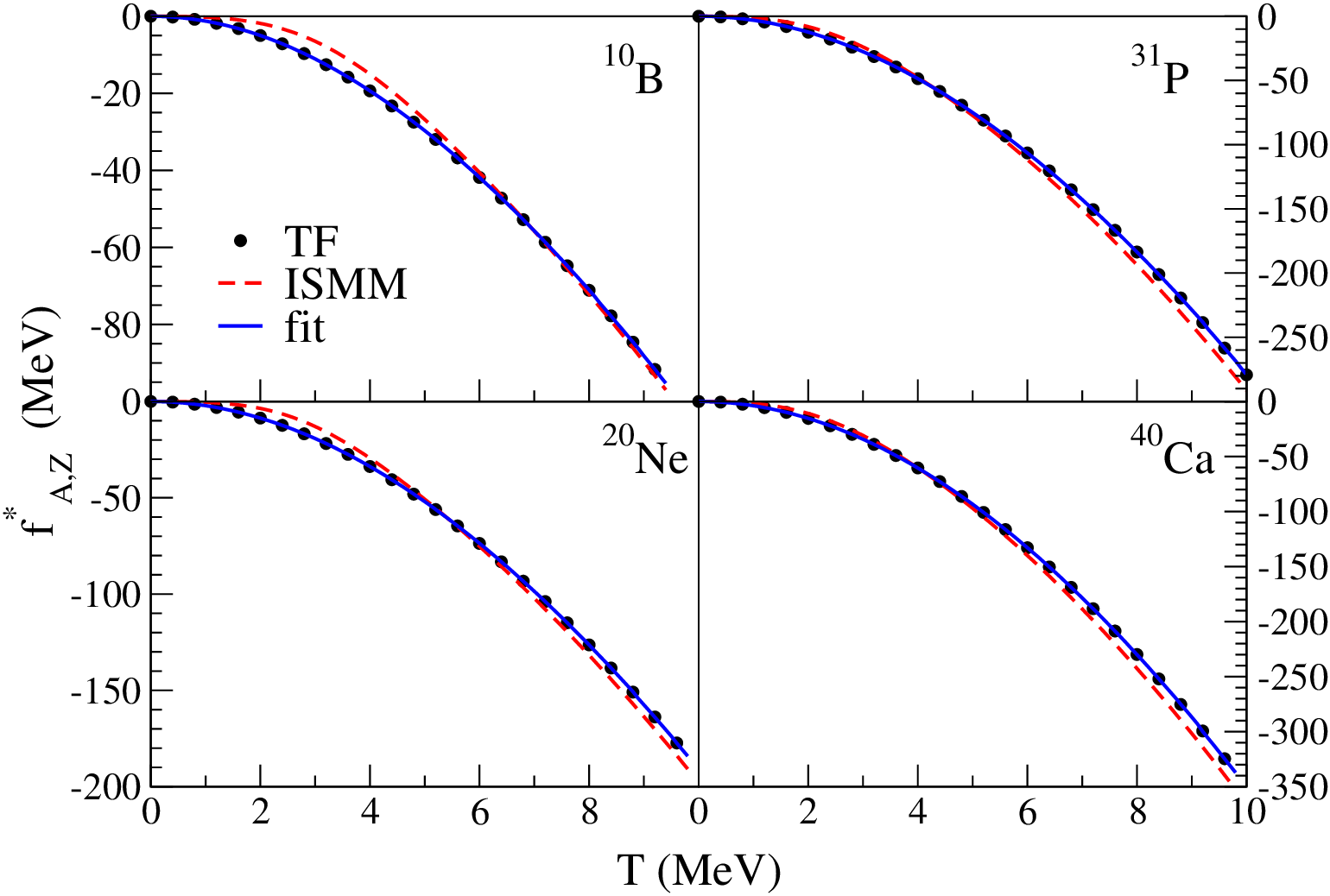}
\caption{\label{fig:feTF}  (Color online) Internal free energy of selected nuclei 
as a function of the temperature.
For details, see the text.
}
\end{figure}

We have also fitted the internal free energies of the nuclei through a simple analytical
formula:

\begin{equation}
f_{A,Z}^*=-T^2\sum_{i=0}^mb_i^{A,Z}T^i\;,
\label{eq:fefit}
\end{equation}

\noindent
where $\{b_i^{A,Z}\}$ are the fit coefficients.
The results are depicted in Fig.\ \ref{fig:feTF} by the full lines, whereas the Thomas-Fermi
calculations are represented by the circles.
As in the previous case, an excellent agreement is obtained with a small number of parameters
($m=5$).
The free energies used in the ISMM are also shown in this picture and are represented by the
dashed lines.
One sees that there are noticeable differences at low temperatures, in the case of the lighter 
nuclei.
Particularly, many more states are suppressed in the ISMM than in the SMM-TF, which 
suggests that the latter should predict larger fragment multiplicities than the former.
This is due to the empirical information on excited states which are taken into account
in the ISMM \cite{ISMMlong}.
In the case of heavier nuclei, the differences are more important at higher temperatures
where the ISMM has more contributions from states in the continuum than the SMM-TF.
However, the determination of the free energy at high temperatures in the ISMM is not as
reliable as in the Thomas-Fermi approximation in the sense that the numerical values of the 
parameters
$\epsilon_0$, $T_c$, and $\beta_0$, used in actual calculations, are not obtained from
a fundamental theory.
They correspond to average values \cite{smm1,smm2} which, sometimes, are
slightly changed by different authors \cite{smm2,ISMMlong,isoscWolfgangBotvina1}.

From the above parametrization to $f_{A,Z}^*$, the entropy and excitation energy
associated with the fragment $(A,Z)$ read:

\begin{equation}
s^*_{A,Z}=2T\sum_{i=0}^mb_i^{A,Z}T^i+T^2\sum_{i=1}^mib_i^{A,Z}T^{i-1}
\label{eq:saztf}
\end{equation}

and

\begin{equation}
\epsilon^*_{A,Z}=T^2\sum_{i=0}^mb_i^{A,Z}T^i+T^3\sum_{i=1}^mib_i^{A,Z}T^{i-1}\;.
\label{eq:eaztf}
\end{equation}

The free energies and equilibrium volumes are calculated using the above expressions for
alpha particles and all nuclei with $A\ge 5$.

\section{\label{sec:results}Results and discussion\protect}
\begin{figure}[tb]
\includegraphics[height=5.5cm]{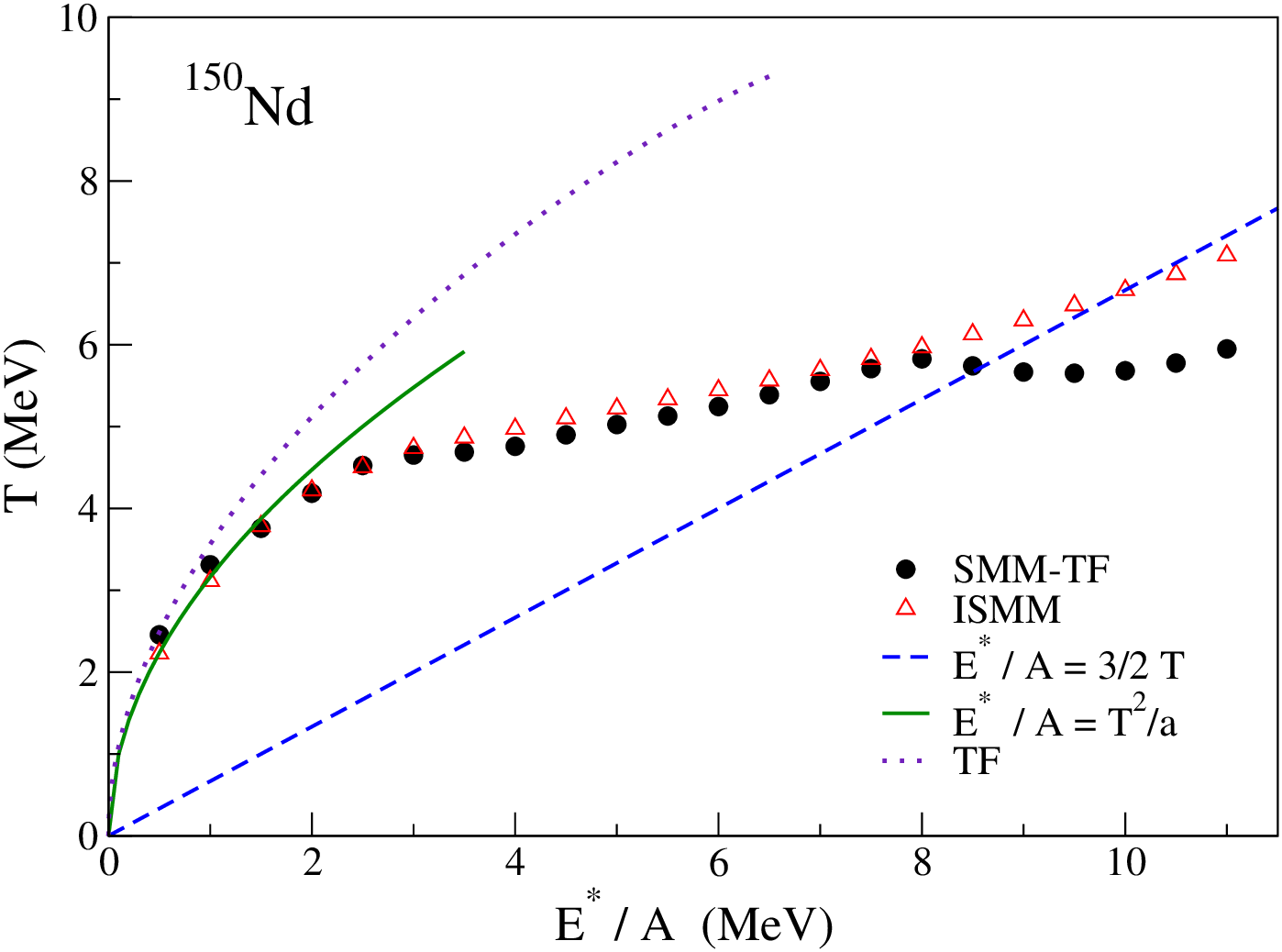}
\caption{\label{fig:cc}  (Color online) Caloric curve associated with the breakup of the
$^{150}$Nd nucleus. The ISMM calculation of Ref.\ \cite{ISMMlong} (triangles) and the SMM-TF
calculation presented in this work (circles) are seen to differ for $E^*/A >8$~MeV.
For reference the excitation energy of the compound nucleus calculated within the
Fermi gas model (full line), the classical gas model (dashed line) as well as the Thomas-Fermi
approach (dotted line) have also been presented. For further details, see the text.}
\end{figure}

The SMM-TF model described in the previous section is now applied to study the breakup of
the $^{150}$Nd nucleus at fixed freeze-out density.
We use $V_\chi/V_0=3$ in all calculations below.
The caloric curve of the system is displayed in Fig.\ \ref{fig:cc}.
Besides the SMM-TF (circles) and the ISMM (triangles) results, the Thomas-Fermi 
calculations for the $^{150}$Nd nucleus are also shown (dotted line), as well as
the Fermi gas (full line) and the Maxwell-Boltzmann (dashed line) expressions.
For $E^*/A\lesssim 8.0$~MeV, both SMM calculations agree fairly well on the 
prediction of the breakup temperatures.
However, a kink in the caloric curve is observed at this point, in the case
of the SMM-TF, indicating that the heat capacity of the system is negative within
a small excitation energy range around this value.
Negative heat capacities have been predicted by many calculations and
have been strongly debated in the recent literature
\cite{smm2,negHC1,negHC2,negHC3,GrossPhysRep,negHC4,negHC5,negHC6,negHC7,negHC8,negHC9}.
However, this feature is normally observed at the onset of the multifragment emission, 
{\it i.e.}
at the beginning of the liquid-gas phase transition \cite{smm2,GrossPhysRep}, whereas
it appears much later in the present calculation.

\begin{figure}[bt]
\includegraphics[height=5.5cm]{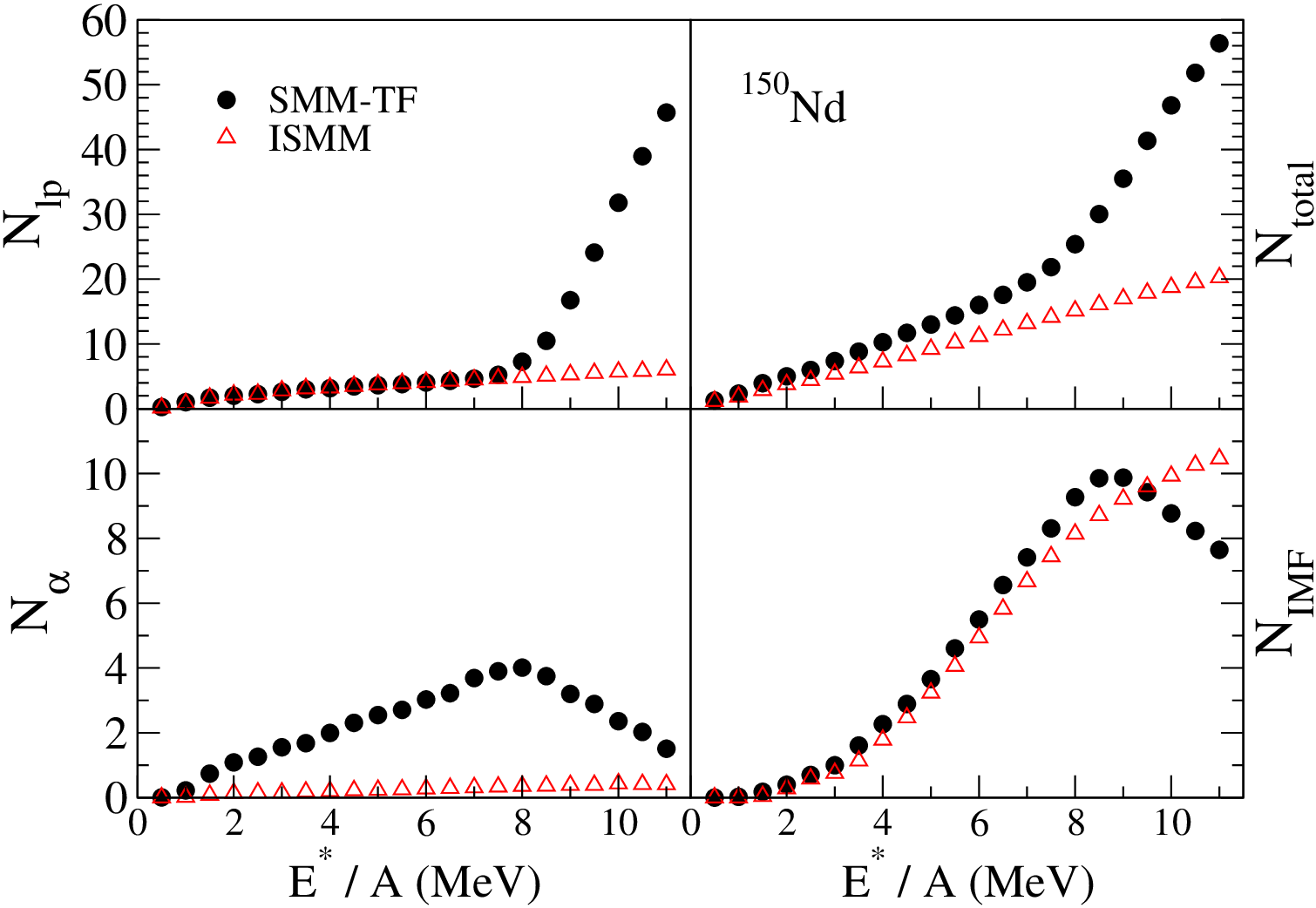}
\caption{\label{fig:multips}  (Color online) Average multiplicity of light particles, alphas,
IMF's and the total fragment multiplicity, as a function of the excitation energy.
For details, see the text.
}
\end{figure}

In order to understand the qualitative differences between the two SMM approaches, we show,
in Fig.\ \ref{fig:multips}, the multiplicity of light particles $N_{lp}$ (all particles
with $A\le 4$, except for alpha particles), the alpha particle $N_\alpha$ and the 
Intermediate Mass Fragment (IMF, $3\le Z\le 15$) $N_{IMF}$ multiplicities, as well as the 
total fragment multiplicity $N_{\rm total}$ as a function of the excitation energy.
It is important to notice that neutrons are included in $N_{lp}$ and $N_{\rm total}$.
One sees that there is a clear disagreement between the two SMM calculations in the 
prediction of the alpha particle multiplicity.
This is due to the construction of the internal free energies in the ISMM \cite{ISMMlong},
which considers empirical low energy discrete states.
Since the first excited state of the alpha particle is around 20~MeV, this strongly 
increases the free energy at low temperatures within the ISMM calculation, 
in contrast to the Thomas-Fermi model calculations.
In the case of the other multiplicities, the agreement between the two model calculations 
is fairly good for excitation energies up to $E^*/A\approx 8$~MeV.
The small discrepancy between $N_{\rm total}$ in the two calculations
can be attributed to the differences in the alpha multiplicities.
All multiplicities rise smoothly up to approximately this excitation energy.
Then, at $E^*/A\approx 8$~MeV, in the SMM-TF calculations, $N_\alpha$ and 
$N_{IMF}$ reach a maximum and decrease from there on.
This behavior is not observed in the case of the ISMM because it takes 
place beyond the energy range considered in the figure.
Another feature also observed in this picture is the sudden change in the slope of the
$N_{\rm total}$ and $N_{lp}$ multiplicities calculated using the SMM-TF model, 
which also takes place at the excitation energy mentioned above, and which is not seen 
in the ISMM results.

Although the Helmholtz free energies of the fragments are somewhat different in both 
calculations, the differences are not large enough to quantitatively explain this 
peculiar behavior, as illustrated in Fig.\ \ref{fig:feTF}.
Therefore, the differences in the multiplicities calculated within the ISMM and
SMM-TF models must be associated with the behavior of the kinetic terms, due to
changes in the free volume in the SMM-TF calculations.

\begin{figure}[b]
\

\

\
\includegraphics[height=5.5cm]{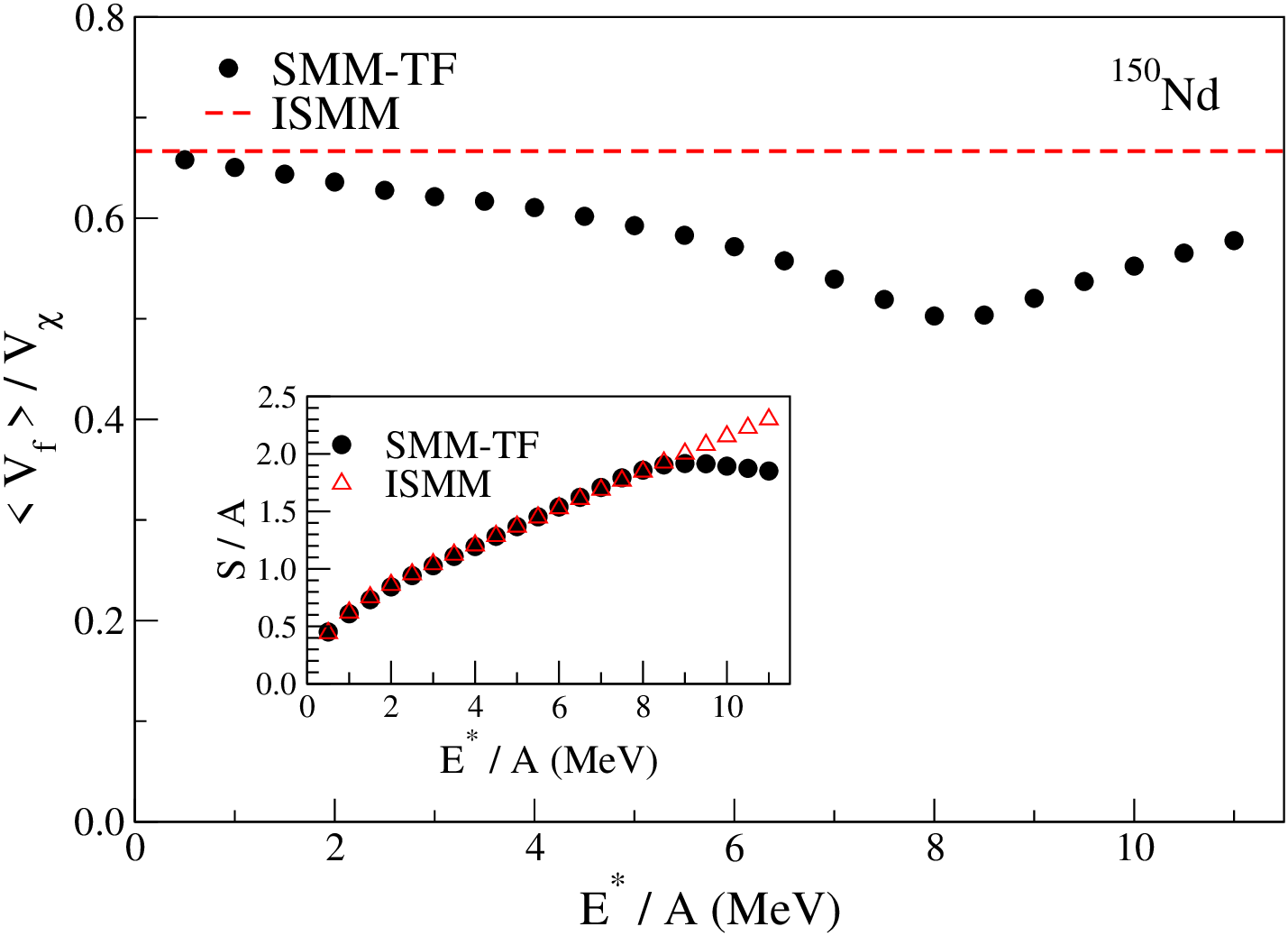}
\caption{\label{fig:vfs}  (Color online) Average free volume and entropy per nucleon 
as a function of the excitation energy calculated within the ISMM and SMM-TF models.
}
\end{figure}

To examine this aspect more closely, we show, in Fig.\ \ref{fig:vfs}, the energy 
dependence of $\langle V_f\rangle$.
It confirms the expectation that $\langle V_f\rangle$ should decrease as $E^*$
increases, owing to the expansion of the fragments' volumes at finite temperature.
However, it reaches a minimum at $E^*/A\approx 8.0$~MeV and rises from this point on.
The logarithmic volume term of the entropy [Eq.\ (\ref{eq:stransfe})] disfavors 
partitions with small free volumes.
Furthermore, the last term in Eq.\ (\ref{eq:stransfe}) also gets larger as $T$ 
increases since, besides being explicitly proportional to $T$, the factor 
$\mid\frac{dV_f}{dT}\mid$ grows faster at high temperatures, as it can be inferred 
from the behavior of the densities shown in Fig.\ \ref{fig:aveRho}.
Therefore, the system favors the emission of very light particles, $N_{lp}$,
which cannot become excited in our treatment, in order to minimize the reduction 
of $V_f$.
Nevertheless, this preference is closely related to the energy conservation 
constraint given by Eq.\ (\ref{eq:econs}).
It is only when the excitation energy becomes sufficiently high that there is 
enough energy for the system to produce a significant number of very light particles.
The inset in Fig.\ \ref{fig:vfs} shows the entropy per nucleon predicted by the 
two SMM treatments.
It reveals that, while in the ISMM case it rises steadily, the entropy saturates,
and even decreases in the SMM-TF model for $8.0 \lesssim E^*/A\lesssim 11.0$~MeV.
The large emission of particles which have no internal degrees of freedom
prevents the entropy from falling off from this point on, since they lead to larger
$\frac{dV_f}{dT}$ (smaller absolute values) by increasing $V_f$, as they do not expand.
One should notice that the reduction of the complex fragment multiplicities does not 
mean that the limiting temperature of the fragments in the different partitions has 
been reached.
In fact, the breakup temperatures obtained in the present calculations are much lower 
than the limiting temperatures of most nuclei, except for the very asymmetric ones, 
as may be seen in the examples given in Fig. \ref{fig:aveRho} and in 
Refs.\ \cite{BLV2,Suraud1987}.
This effect on the fragments produced should appear at much higher excitation energies,
as those fragments have excitation energies much smaller than the original nucleus,
since an appreciable amount of energy is used in the breakup of the system.
Therefore, the back bending of the caloric curve and the small plateau observed in 
Fig.\ \ref{fig:cc} are strongly ruled by the changes in the free volume.
As a consequence of this fact, the phase transition at high excitation energy takes 
place at approximately constant entropy.

\begin{figure}[bt]
\includegraphics[height=5.5cm]{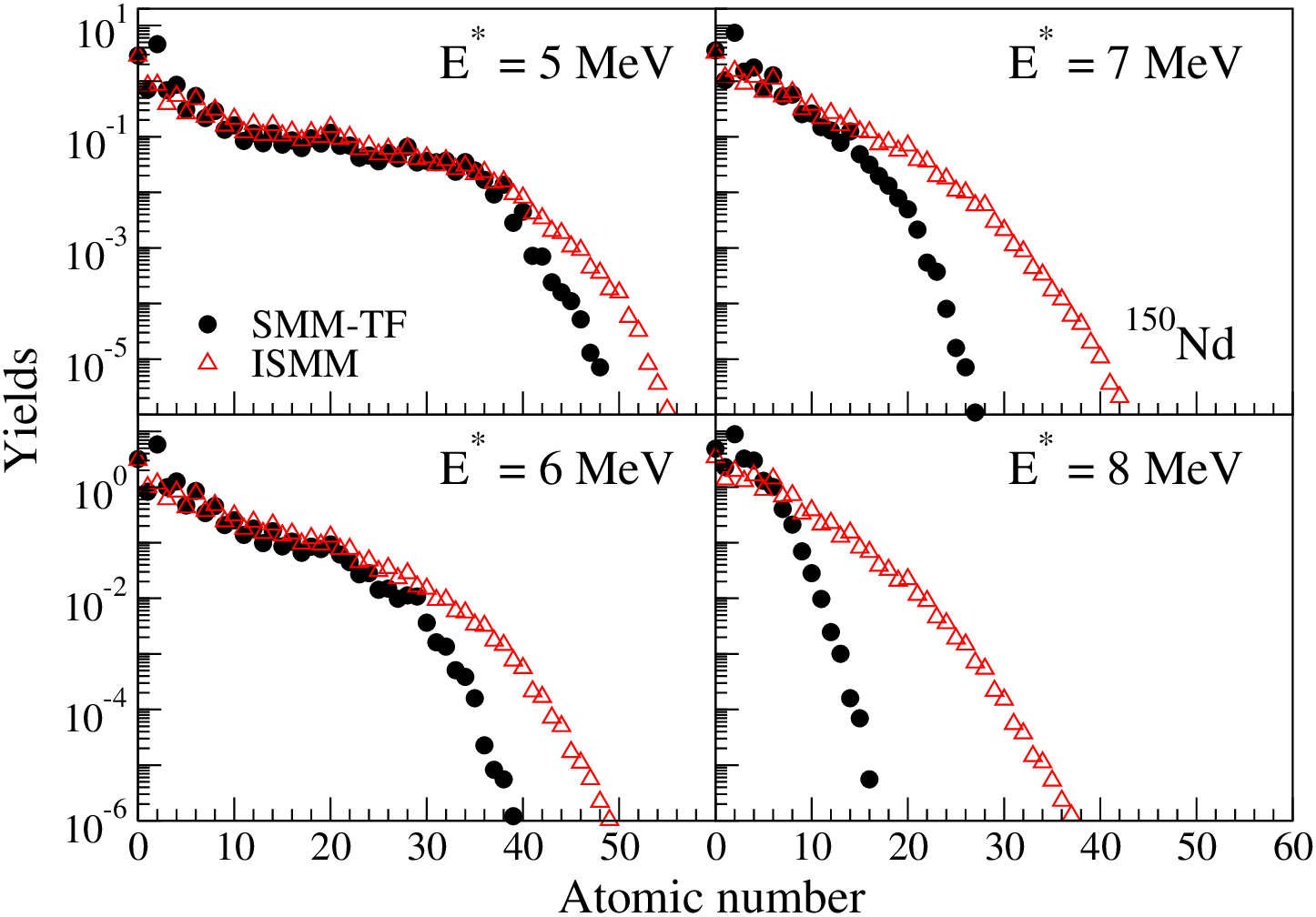}
\caption{\label{fig:yz}  (Color online) Charge distribution in the breakup of 
the $^{150}$Nd nucleus at four different excitation energies.
}
\end{figure}

This observation is also corroborated by the charge distributions shown 
in Fig.\ \ref{fig:yz} for four different excitation energies: $E^*/A=5$, 6, 7 and 8~MeV.
It shows that the multiplicity of heavy fragments is strongly reduced in the SMM-TF 
calculations as the excitation energy increases, although they are not completely 
ruled out of the possible fragmentation modes.
In particular, the SMM-TF model systematically gives much lighter fragments than 
the ISMM, for the reasons just discussed.

\begin{figure}[bt]
\includegraphics[height=5.5cm]{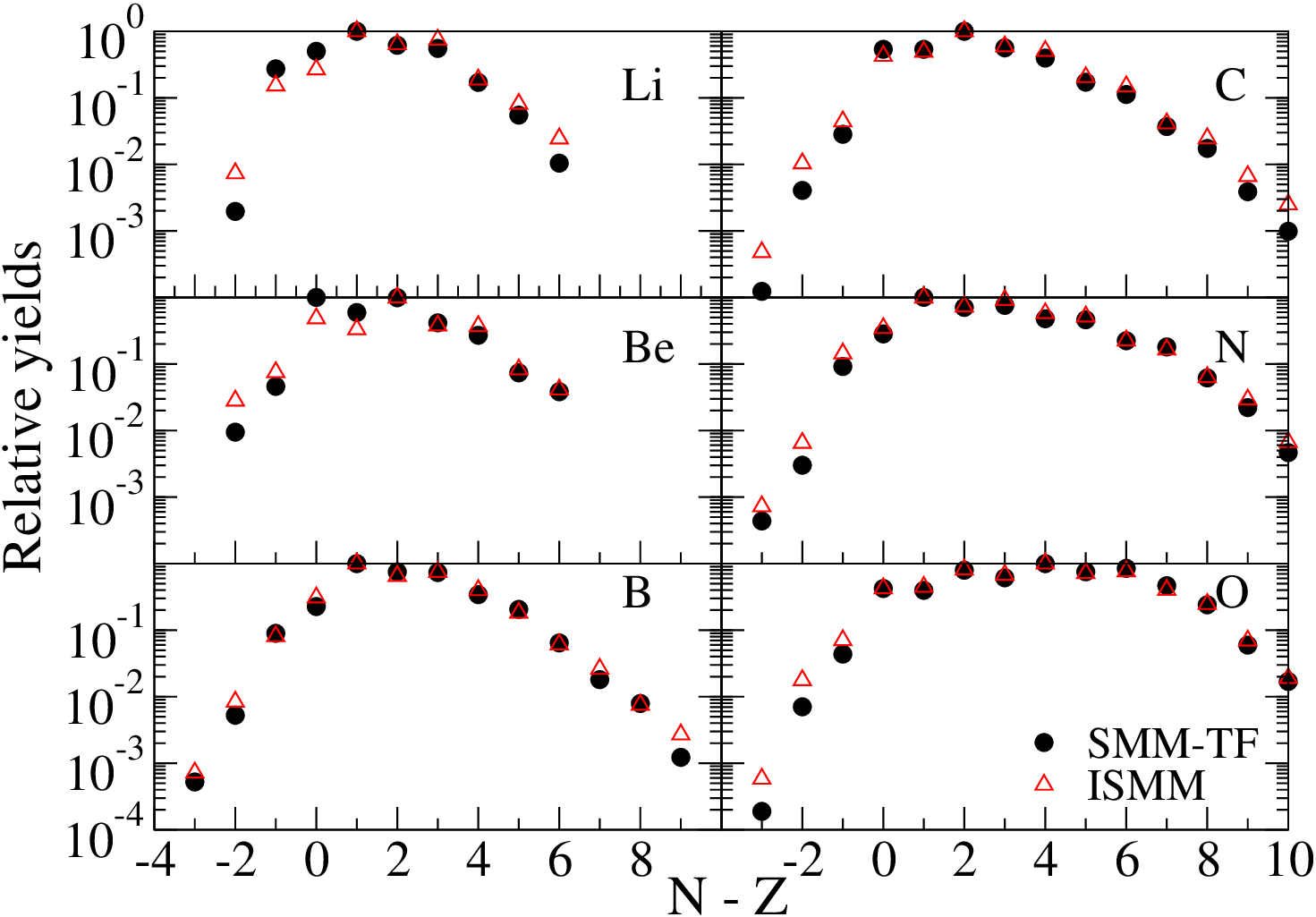}
\caption{\label{fig:isodist}  (Color online) Isotopic distribution of selected 
nuclear species produced in the breakup of the $^{150}$Nd nucleus at $E^*/A=6.0$~MeV.
}
\end{figure}

Even though the fragments are not directly affected by their limiting temperatures 
at the excitation energies we consider, the reduction of the entropy associated with 
the volume affects the fragment species in different ways.
Indeed, since the proton rich nuclei tend to be more unstable, they are hindered 
due to these dilatation effects more strongly than the other isotopes.
Owing to their larger volumes at a given temperature $T$, partitions containing 
proton rich fragments have smaller entropies than the others.
Therefore, one should expect to observe a reduction in the yields of these fragments.
This qualitative reasoning is confirmed by the results presented in 
Fig.\ \ref{fig:isodist}, which displays the isotopic distribution of some selected 
light fragments, produced at $E^*/A=6.0$~MeV.
One sees that, even though both SMM models make similar predictions for many 
observables at this excitation energy, the role played by the free volume effects 
just discussed is non-negligible.
Since the limiting temperatures, as well as the equilibrium density at temperature 
$T$, is sensitive to the effective interaction used \cite{BLV2,Suraud1987}, these 
findings suggest that careful comparisons with experimental data may provide valuable 
information on the EOS.

\section{\label{sec:conclusions}Concluding remarks\protect}
We have modified the SMM to incorporate the Helmholtz free energies and equilibrium 
densities of nuclei at finite temperature from the results obtained with the 
Thomas-Fermi approximation using Skyrme effective interactions.
Owing to the reduction of the fragments' translational energy at finite temperature,
the model predicts the existence of two temperatures associated with the same total energy.
This feature is directly associated with the reduction of the free volume due to the
expansion of the fragments' volumes.
If this statistical treatment proves to be more appropriate to describe the nuclear
multifragmentation process than its standard version, the determination of the isotopic
temperatures, at high excitation energies, should be carefully reexamined, since one
tacitly assumes a univocal relationship between the temperature and the excitation energy
in the derivation that leads to the corresponding formulae \cite{Albergo}.

The thermal dilatation of the fragments' volumes also has important 
consequences on the fragmentation modes.
For excitation energies larger than approximately 8 MeV per nucleon, it favors enhanced
emission of particles which have no internal degrees of freedom (very light nuclei, 
protons and neutrons), leading to the onset of a gas transition at excitation energies 
around this value.
The existence of a small kink in the caloric curve, as well as a plateau, for a system
at constant volume is qualitatively different from the results obtained in previous SMM 
calculations where these features were observed only at (or at least at nearly) constant 
pressure \cite{smmIsobaric}.

Since many-particle multiplicities, such as those associated with the IMF's and the 
light particles, are very different in both statistical treatments for excitation energies 
larger than 8~MeV per nucleon, we believe that careful comparisons with experimental data 
may help to establish which treatment is better suited for describing the multifragment 
emission.
Furthermore, since the isotopic distribution turns out to be sensitive to the treatment 
even at lower excitation energies, this suggests that one may obtain important information 
on the EOS by using different Skyrme effective interactions in the SMM-TF calculations.
Particularly, this modified SMM model is appropriate to investigate the density dependence 
of the symmetry energy recently discussed 
\cite{isoscalingIndraGSI2005,isoSymmetryBotvina2006,symEnergyShatty2007,isoMassFormula2008}.

\begin{acknowledgments}
We would like to acknowledge CNPq, FAPERJ, and the PRONEX program under contract 
No E-26/171.528/2006, for partial financial support.
This work was supported in part by the National Science Foundation under Grant
Nos.\ PHY-0606007 and INT-0228058.
AWS is supported by Joint Institute for Nuclear Astrophysics at MSU
under NSF-PFC grant PHY 02-16783.
\end{acknowledgments}

\appendix

\section{\label{sec:ThomasFermi}The finite temperature Thomas-Fermi approximation}
The Thomas-Fermi approximation to nuclear systems is thoroughly discussed
in Refs.\ \cite{TFBrack1,TFBrack2,SuraudVautherin,Suraud1987}.
Thus, we review its essential features below in order to give a full account
of all calculations presented in this work.

The equilibrium configuration of a nucleus at temperature $T$ is found by minimizing the
thermodynamical potential $\Omega$ with respect to the number density $\rho_\alpha$
($\alpha=p,n$ for protons or neutrons):

\begin{equation}
\Omega={\cal F}[\rho]-\sum_\alpha\int\;d^3\vec{r}\;\mu_\alpha\rho_\alpha
\label{eq:mintp}
\end{equation}

\noindent
where the Helmholtz free energy is given by

\begin{equation}
{\cal F}[\rho]=\int\left[{\cal H}_{\rm nucl}
+{\cal H}_{\rm Coul}-T\sum_\alpha{\cal S}_\alpha\right]d^3\vec{r}\;.
\label{eq:fetf}
\end{equation}

\noindent
In the above expression,  ${\cal S}_\alpha$ denotes the entropy density associated with the
species $\alpha$, $\mu_\alpha$ is the corresponding chemical potential,
$\cal{H}_{\rm nucl}$ is the nuclear energy density of the system, and the Coulomb term reads:

\begin{eqnarray}
{\cal H}_{\rm Coul}&=&\frac{e^2}{2}\rho_p(\vec{r})
                    \int\;\frac{\rho_p(\vec{r'})}{\mid\vec{r}-\vec{r'}\mid}\;d^3\vec{r'}
\nonumber\\
                   &-&\frac{3}{4}e^2\left(\frac{3}{\pi}\right)^{1/3}\rho_p^{4/3}(\vec{r})\;.
\label{eq:hCoul}
\end{eqnarray}

\noindent
The second term above corresponds to an approximation to the exchange contribution to the
Coulomb energy \cite{VautherinNegeleDensityMatrix,CoulombExchange}.

The expression for $\cal{H}_{\rm nucl}$ given in Ref.\ \cite{TFBrack1} may be rewritten as

\begin{equation}
{\cal H}_{\rm nucl}={\cal H}_0+{\cal H}_\tau+{\cal H}_{\rm grad}+{\cal H}_J
\label{eq:H}
\end{equation}

\noindent
where

\begin{eqnarray}
{\cal H}_0&=&\frac{t_0}{2}\left[(1+\frac{x_0}{2})\rho^2-(x_0+\frac{1}{2})(\rho_n^2+\rho_p^2)
           \right]\\
         &+&\frac{t_3}{12}\rho^\sigma\left[(1+\frac{x_3}{2})\rho^2
           -(x_3+\frac{1}{2})(\rho_n^2+\rho_p^2)\right]\,,\nonumber
\label{eq:h0}
\end{eqnarray}

\begin{equation}
{\cal H}_\tau=\frac{\hbar^2}{2m^*_p}\tau_p+\frac{\hbar^2}{2m^*_n}\tau_n\;,
\label{eq:htau}
\end{equation}

\begin{eqnarray}
\label{eq:hgrad}
{\cal H}_{\rm grad}&=&\frac{1}{64}\left[9t_1-5t_2(1+\frac{4}{5}x_2)\right]
                    \left(\vec{\bigtriangledown}\rho\right)^2\\
                   &-&\frac{1}{64}\left[3t_1(1+2x_1)+t_2(1+2x_2)\right]
                    \left(\vec{\bigtriangledown}\rho_n-
                    \vec{\bigtriangledown}\rho_p\right)^2\;,\nonumber
\end{eqnarray}

\begin{equation}
{\cal H}_J=\frac{1}{2}W_0\left[\vec{J}\cdot\vec{\bigtriangledown}\rho
         +\vec{J}_n\cdot\vec{\bigtriangledown}\rho_n
         +\vec{J}_p\cdot\vec{\bigtriangledown}\rho_p\right]\;,
\label{eq:hj}
\end{equation}

\noindent
the total density is denoted by $\rho=\rho_n+\rho_p$ and $\vec{J}=\vec{J}_p+\vec{J}_n$ is
the spin-orbit density.
The kinetic factor $\tau_\alpha$ is given by

\begin{equation}
\tau_\alpha=\frac{1}{2\pi^2}\left(\frac{2m^*_\alpha}{\hbar^2}\right)^{5/2}T^{5/2}
I_{3/2}(y_\alpha)
\label{eq:tau}
\end{equation}

\noindent
where

\begin{eqnarray}
\frac{\hbar^2}{2m^*_\alpha}&=&\frac{\partial}{\partial \tau_\alpha}{\cal H}_{\rm nucl}
\nonumber\\
&=&\frac{\hbar^2}{2m}+\frac{1}{8}\left[t_1(1-x_1)+3t_2(1+x_2)\right]\rho_\alpha
\nonumber\\
  &+&\frac{1}{4}\left[t_1(1+\frac{x_1}{2})+t_2(1+\frac{x_2}{2})\right]\rho_{\alpha'}
\label{eq:effectm}
\end{eqnarray}

\noindent
and $\rho_{\alpha'}=\rho_p\,(\rho_n)$ if $\alpha=n\,(p)$.
The Fermi-Dirac integral

\begin{equation}
I_{n/2}(y)=\int_0^\infty\;dx\;\frac{x^{n/2}}{1+\exp(x-y)}
\label{eq:FermiI}
\end{equation}

\noindent
is efficiently calculated using the formulae given in Ref.\ \cite{FermiIntegrals}, where
one also finds approximations to the inverse function $y(I_{n/2})$.
The latter is determined from the number density

\begin{equation}
\rho_\alpha=\frac{1}{2\pi^2}
\left(\frac{2m^*_\alpha}{\hbar^2}\right)^{3/2}T^{3/2}I_{1/2}(y_\alpha)
\;.
\label{eq:rhoy}
\end{equation}

\noindent
The entropy density ${\cal S}_\alpha$ can then be easily calculated

\begin{equation}
{\cal S}_\alpha=\frac{5}{3}
\frac{\hbar^2}{2m^*_\alpha}\frac{\tau_\alpha}{T}-\rho_\alpha y_\alpha
\;.
\label{eq:sdens}
\end{equation}

\noindent
The parameter set $\{x_i,t_i,\sigma,W_0\}$, $i=0,1,2,3$, for the Skyrme
interaction used in this work, SKM, is listed in Ref.\ \cite{TFBrack1}.
Since we stay in the zero{\it-th} order approximation in $\hbar$, 
$\vec{J}_\alpha=0$ and then ${\cal H}_J$ does not contribute to $\cal{H}_{\rm nucl}$
\cite{TFBrack1}.

Following Suraud and Vautherin \cite{SuraudVautherin,Suraud1987}, the equilibrium 
configuration is found by iterating the densities at the {\it k}-th step according to

\begin{equation}
\rho^{(k+1)}_\alpha=\rho^{(k)}_\alpha\left[1-\lambda\left(B^{(k)}_\alpha-\mu^{(k)}_\alpha\right)
\right]
\label{eq:rhoiter}
\end{equation}

\noindent
where

\begin{equation}
\mu^{(k)}_\alpha=\frac{1}{N_\alpha}\int\;d^{3}\vec{r}\;B^{(k)}_\alpha(\vec{r})
                 \rho^{(k)}_\alpha(\vec{r}),
\label{eq:mu}
\end{equation}

\

\noindent
$N_p=Z$, $N_n=A-Z$, and

\begin{equation}
B^{(k)}_\alpha=\frac{\delta {\cal F}}{\delta \rho^{(k)}_\alpha}\;.
\label{eq:b}
\end{equation}

\noindent
The parameter $\lambda$ is chosen to be small enough in order to ensure that the
first order approximation given by Eq.\ (\ref{eq:rhoiter}) remains valid.

In our numerical implementation, we have assumed spherical symmetry, and discretized 
the space using a mesh spacing $\Delta R = 0.1$~fm, which suffices for our purposes.
As suggested in Refs.\ \cite{SuraudVautherin,Suraud1987}, the second term of 
Eq.\ (\ref{eq:hgrad}) is neglected since it is small and may lead to numerical instabilities.
Similarly to the treatment adopted in Ref.\ \cite{SuraudVautherin}, the gradient density 
terms are calculated at the mesh point $r_{i+1/2}=(i+1/2)\Delta R$, using \cite{Koonin}

\begin{equation}
\frac{\partial}{\partial r}\rho(r_{i+1/2})=\frac{\rho(r_i)-\rho(r_{i-1})}{2(\Delta R/2)}
+{\cal O}[(\Delta R)^2]\;,
\label{eq:disGrad}
\end{equation}

\noindent
which turned out to be numerically stable.

Due to the important contributions associated with unbound states at high temperatures,
the above treatment is not accurate for $T\gtrsim 4$~MeV, as pointed out by Bonche, Levit,
and Vautherin \cite{BLV1}.
Therefore, these authors have proposed a method to extend the Hartree-Fock calculations 
to higher temperatures.
As they have noticed, there are two solutions of the Hartree-Fock equations for a given 
chemical potential.
One of them corresponds to a nucleus in equilibrium with its evaporated particles whereas
the other is associated with the nucleon gas.
Thus, in their formalism, the properties of the hot nucleus is obtained by subtracting
the thermodynamical potential associated with an introduced nucleon gas $\Omega_G$ from that
corresponding to the nucleus in equilibrium with its evaporated gas $\Omega_{NG}$.
Except for the Coulomb energy, there is no interaction between the gas and the
nucleus-gas system.

This approach has been successfully applied by these authors \cite{BLV1,BLV2} and has 
been adapted to the finite temperature Thomas-Fermi approximation by Suraud \cite{Suraud1987}.
More precisely, the thermodynamical potential associated with the nucleus is given by

\begin{equation}
\Omega_N=\Omega_{NG}-\Omega_G+E_{\rm Coul}\;.
\label{eq:OmegaN}
\end{equation}

\noindent
One should notice that, by construction, $\Omega_{NG}$ and $\Omega_G$ do not contain any
Coulomb contribution.
More specifically, one defines the subtracted free energy

\begin{eqnarray}
{\cal F}^{\rm sub}&=&\int\;\left[{\cal H}_{\rm nucl}^{NG}-{\cal H}_{\rm nucl}^{G}
-T\sum_\alpha\left({\cal S}_\alpha^{NG}-{\cal S}_\alpha^G\right)\right] \;d^3\vec{r}\nonumber
\\
&+&\int\;{\cal H}_{\rm Coul}^{\rm sub}\;d^3\vec{r}
\label{eq:fengg}
\end{eqnarray}

\noindent
where the subtracted Coulomb energy density, in the last term of this expression, reads

\begin{eqnarray}
\label{eq:Coulngg}
{\cal H}_{\rm Coul}^{\rm sub}&\equiv&\frac{e^2}{2}\rho_p(\vec{r})\int\;d^3\vec{r'}
\frac{\rho_p(\vec{r'})}{\mid \vec{r}-\vec{r'}\mid}\\\
&-&\frac{3}{4}e^2\left(\frac{3}{\pi}\right)^{1/3}
\left[\left(\rho^{NG}_p\right)^{4/3}
-\left(\rho^{G}_p\right)^{4/3}\right]\;,\nonumber
\end{eqnarray}

\noindent
and the subtracted density $\rho_p$:

\begin{equation}
\rho_p(\vec{r})=\rho_p^{NG}(\vec{r})-\rho_p^G(\vec{r})
\label{eq:rhosub}
\end{equation}

\noindent
is the quantity that enters in the direct part of the Coulomb energy.

The iteration scheme given by Eq.\ (\ref{eq:rhoiter}) remains unchanged if one rewrites
$B_\alpha^{(k)}$ as

\begin{equation}
B_\alpha^{(k,\gamma)}=
\pm\frac{\delta {\cal F}^{\rm sub}}{\delta \rho^{(k,\gamma)}_\alpha}\;,
\label{eq:bnew}
\end{equation}

\noindent
where the super-index $(k,\gamma)$ denotes the quantity associated with the gas ($\gamma=G$)
or the nucleus-gas ($\gamma=NG$) at the {\it k-}th stage of the iteration.
The positive sign is associated with the $NG$ solution whereas the negative sign is 
used in the other case.
The proton and neutron chemical potentials are given by an expression similar to
Eq.\ (\ref{eq:mu})

\begin{eqnarray}
\mu^{(k)}_\alpha=\frac{1}{N_\alpha}\int\;d^{3}\vec{r}\;& \Big\{ &
                   B^{(k,NG)}_\alpha(\vec{r})\rho^{(k,NG)}_\alpha(\vec{r})\nonumber\\
                &-&B^{(k,G)} _\alpha(\vec{r})\rho^{(k,G)}_\alpha(\vec{r})\Big\},
\label{eq:muNew}
\end{eqnarray}

\noindent
since $\rho^{NG}$ and $\rho^G$ are constrained by

\begin{equation}
N_\alpha=\int\;d^3\vec{r}\;
\left[\rho^{(k,NG)}_\alpha(\vec{r})-\rho^{(k,G)}_\alpha(\vec{r})\right]
\;.
\label{eq:nrho}
\end{equation}

One then starts with a reasonable guess for $\rho_{\alpha}^{NG}$ and $\rho_\alpha^{G}$, 
which can be a Woods-Saxon density for the former and a small constant value for the 
latter (subject to the condition $\rho_\alpha>0$), obeying the constraint given by 
the above expression, and apply the iteration scheme just described.
Ideally, convergence is reached when $B^{(k,\gamma)}_\alpha(\vec{r})-\mu_\alpha^{(k)}$
vanishes, so that $\rho_\alpha^{(k,\gamma)}$ becomes stationary.
In practice, one can monitor the quantity \cite{Suraud1987}

\begin{eqnarray}
\Delta E_\alpha^2=\int\;d^3\vec{r}\; &\Big\{ &
      \left(B^{(k,NG)}_\alpha(\vec{r})-\mu_\alpha^{(k)}\right)^2\rho_\alpha^{(k,NG)}
\nonumber\\
     &+& \left(B^{(k,G)}_\alpha(\vec{r})-\mu_\alpha^{(k)}\right)^2\rho_\alpha^{(k,G)}\Big\}
\label{eq:tol}
\end{eqnarray}

\noindent
and stop the iteration when the established tolerance is reached.
The Helmholtz free energy of the nucleus can then be easily calculated through 
Eq. (\ref{eq:fengg}), so that the internal free energy of the nucleus is

\begin{equation}
f_{A,Z}^*(T)={\cal F}^{\rm sub}(T)-{\cal F}^{\rm sub}(T=0)\;.
\label{eq:feaztf}
\end{equation}

We have used the approximation just described in this Appendix to calculate $f_{A,Z}^*$
for all the fragments entering in the SMM, with $A\ge 5$ (and alpha particles) 
from $T=0$~MeV up to the limiting temperature \cite{BLV1,BLV2,Suraud1987} in steps of 0.1~MeV.

\bibliography{smmtf}

\end{document}